\documentclass{ws-ijmpb}

\usepackage{graphicx}
\usepackage{dcolumn}
\usepackage{bm}

\begin{document}

\markboth{T. Ihn, C. Ellenberger, C. Yannouleas, U. Landman, 
K. Ensslin, D. Driscoll, and A.C. Gossard}
{Quantum dots based on parabolic quantum wells: importance of electronic correlations}

%
\catchline{}{}{}{}{}
%

\title{QUANTUM DOTS BASED ON PARABOLIC QUANTUM WELLS:\\
IMPORTANCE OF ELECTRONIC CORRELATIONS}

\author{THOMAS IHN, CHRISTOPH ELLENBERGER, KLAUS ENSSLIN}

\address{Solid State Physics Laboratory, ETH Zurich, CH-8093 Zurich, Switzerland,
ihn@phys.ethz.ch}

\author{CONSTANTINE YANNOULEAS, UZI LANDMAN}

\address{School of Physics, Georgia Institute of Technology,
Atlanta, Georgia 30332-0430}

\author{DAN C. DRISCOLL, ART C. GOSSARD}

\address{Materials Department, University of California,
Santa Barbara, CA 93106}

\maketitle

\begin{history}
\received{Day Month Year}
\revised{Day Month Year}
\end{history}

\begin{abstract}
We present measurements and theoretical interpretation of the magnetic field dependent excitation spectra of a two-electron quantum dot. The quantum dot is based on an Al$_x$Ga$_{1-x}$As parabolic quantum well with effective $g^\star$-factor close to zero. Results of tunneling spectroscopy of the four lowest states are compared to exact diagonalization calculations and a generalized Heitler--London approximation and good agreement is found. Electronic correlations, associated with the formation of an H$_2$-type Wigner molecule, turn out to be very important in this system.
\end{abstract}


\section{Introduction}
The attempt to understand the effects of Coulomb interaction in solids---with exchange- and correlation effects being the most interesting manifestations---has always been a driving force behind research in solid state systems. Nowadays, fabrication techniques for quantum dots offer the unique possibility to tailor interacting quantum systems to an unprecedented degree. Understanding quantum dot helium, a man-made two-electron system, is a hallmark for our ability to design and control more complex interacting quantum systems such as those required for the implementation of quantum information processing schemes. Here, we present a study of quantum dot Helium fabricated on a parabolic quantum well\cite{Ellenberger06} which goes beyond earlier work\cite{Kouwenhoven97} by detecting higher lying excited states at high magnetic fields and by presenting evidence for the importance of correlation effects.

\section{Parabolic Quantum Wells}
\begin{figure}[b]
\parbox[b]{6cm}{
\includegraphics[width=6cm]{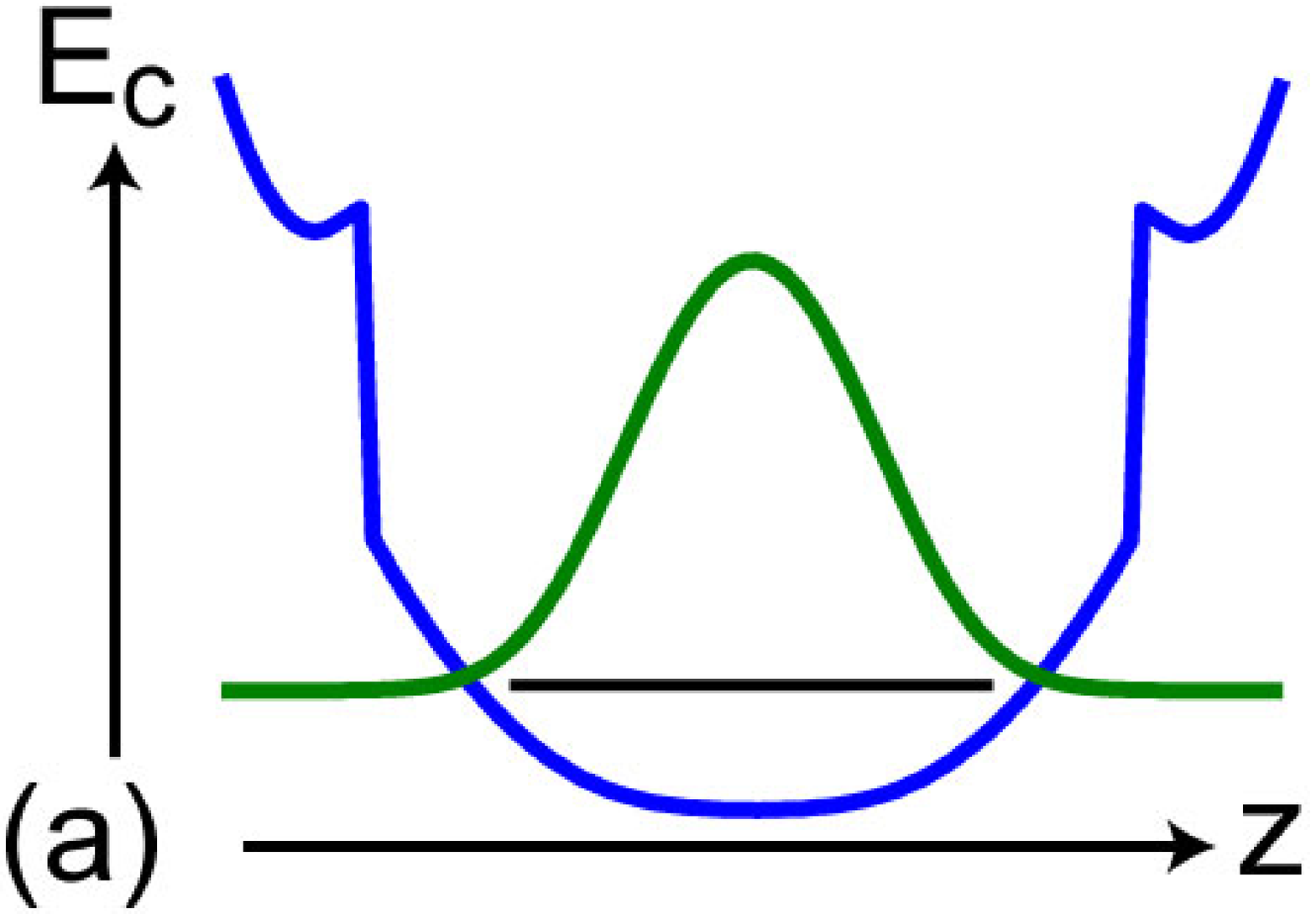}}\hfill
\parbox[b]{6cm}{
\includegraphics[width=6cm]{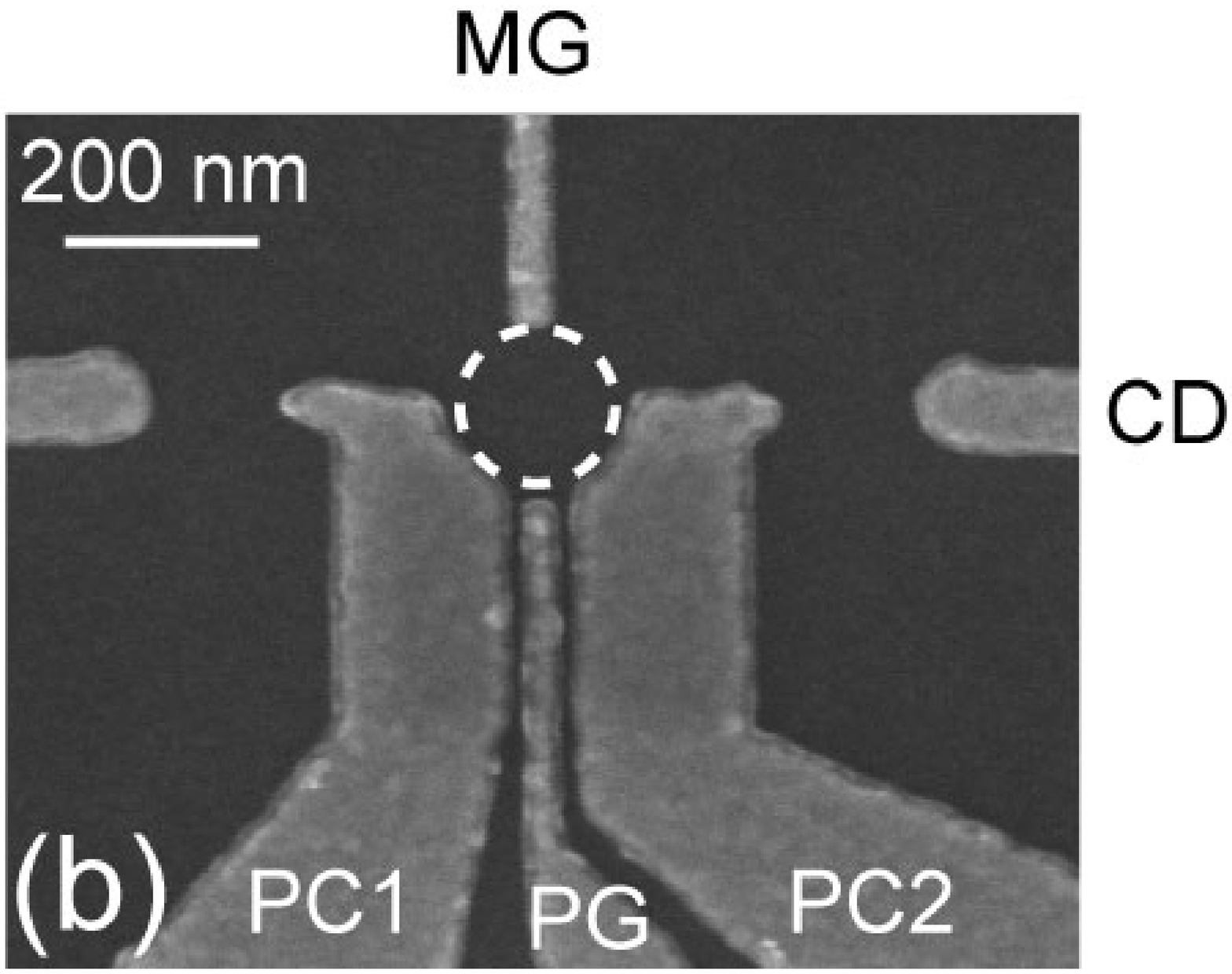}}
\caption{(a) Schematic plot of the conduction band edge in a parabolic quantum well as a function of coordinate $z$ measured in the direction of crystal growth. The lowest bound
subband state with its envelope wave function is indicated. (b) Scanning electron micrograph of the few-electron quantum dot with integrated charge readout fabricated on a PQW.}
\label{fig1}
\end{figure}

Parabolic quantum wells (PQWs) based on GaAs/Al$_x$Ga$_{1-x}$As heterostructures have been introduced in 1988 for creating high mobility quasi-three-dimensional electron gases.\cite{Shayegan88} Within the effective mass approximation their conduction and valence band edges vary parabolically along the growth direction, as schematically depicted in Fig.~\ref{fig1}(a), caused by the varying Al content $x$ of the Al$_x$Ga$_{1-x}$As material. For example, the conduction band edge is described by
\begin{equation}
E_\mathrm{c}(z) = \frac{1}{2}az^2,
\label{eq1}
\end{equation}
where $a$ is the curvature parameter of the parabolic confinement.

A PQW can be filled with electrons using standard remote doping techniques. The electrons occupy subbands which form in the well self-consistently under the influence of the parabolic confinement and screening.
With increasing two-dimensional electron density, this interplay leads---in the limit of large density---to a constant three-dimensional electron density $n_\mathrm{3D}=\varepsilon\varepsilon_0a/e$ along with an increasing width of the density distribution in growth direction. Here, $\varepsilon$ is the relative dielectric constant of the material.

A very peculiar property of PQWs is the rigidity of the subband wave functions under the influence of an external homogeneous electric field $E_z$ in growth direction, given that the two-dimensional electron density is constant. Such a field can, for example, be created with metallic gates on top and at the back side of the substrate. It is equivalent to an electrostatic potential changing linearly with $z$ which adds to the built-in $z^2$-confinement in Eq.~(\ref{eq1}) and leads to a net parabolic potential with the same curvature parameter $a$, but shifted by $\Delta z=eE_z/a$ in growth direction. If this shifted potential is populated with the same two-dimensional electron density, it will develop the same self-consistent subbands with the same envelope functions, but shifted in space by $\Delta z$. This property of PQWs has been exploited for mapping the subband wave function in real space.\cite{Salis97}

In the Al$_x$Ga$_{1-x}$As material, the effective mass $m^\star$ and the effective $g$-factor $g^\star$ depend on the Al fraction $x$. For example, the effective $g$-factor varies according to
\begin{equation}
g^\star = -0.44 + 3.833 x \;\;\; \mbox{for $0\leq x\leq 0.3$}.
\label{eq2}
\end{equation}
The wave functions of subbands in PQWs probe regions of varying $x$ and therefore feel an average $g^\star$. This average varies, when the wave function is shifted, providing a gate voltage tunable $g^\star$, i.e., a gate voltage tunable Zeeman splitting in a magnetic field.\cite{Salis01} Furthermore, PQWs can be designed to have certain $g^\star$ values when the wave function is positioned symmetrically around the minimum of the parabola at zero gate voltages by deliberately choosing a certain offset value for $x$ at the well center. For example, PQWs with $g^\star\approx 0$ can be grown.

Recently, PQWs have served as the basic material for the fabrication of mesoscopic devices such as quantum point contacts\cite{Salis99} and quantum dots.\cite{Ellenberger06,Lindemann02,Lindemann02a} In particular, the spin states in a quantum dot fabricated on a PQW were investigated\cite{Lindemann02a} and few-electron dots with 0--3 electrons were realized.\cite{Ellenberger06} Such few-electron dots are candidates for the implementation of spin-qubits for quantum information processing.\cite{Loss98}
Dots on PQWs promise the additional feature of a tunable $g^\star$ which could be used for the realization of gate-controlled single-qubit operations.\cite{Cerletti05}

\section{Sample and Experimental Setup}

The sample used for this study\cite{Ellenberger06} is a quantum dot (QD) with integrated charge readout fabricated on a PQW by electron-beam lithography. The PQW material has been designed and measured to have $g^\star\approx 0$. A highly doped layer, 1.3~$\mu$m below the well serves as a back gate. A scanning electron micrograph of the sample is shown in Fig.~\ref{fig1}(b). The QD is formed by applying negative voltages to the gates PC1, PC2, PG and MG.
The quantum point contact formed between gates PC2 and CD has been used as a detector for the charge in the quantum dot\cite{Field93} and allows to count the number of electrons in the dot from 0 upwards. Experiments were performed in a dilution refrigerator with a base temperature of 100~mK using standard low-frequency DC and AC conductance measurement techniques.

\section{Excited State Spectroscopy of Quantum Dot Helium}

The differential conductance of the QD has been measured in the few-electron regime as a function of plunger gate voltage $V_\mathrm{pg}$ and source--drain voltage $V_\mathrm{bias}$.
\begin{figure}[bthp]
\parbox[b]{6cm}{
\includegraphics[width=6cm]{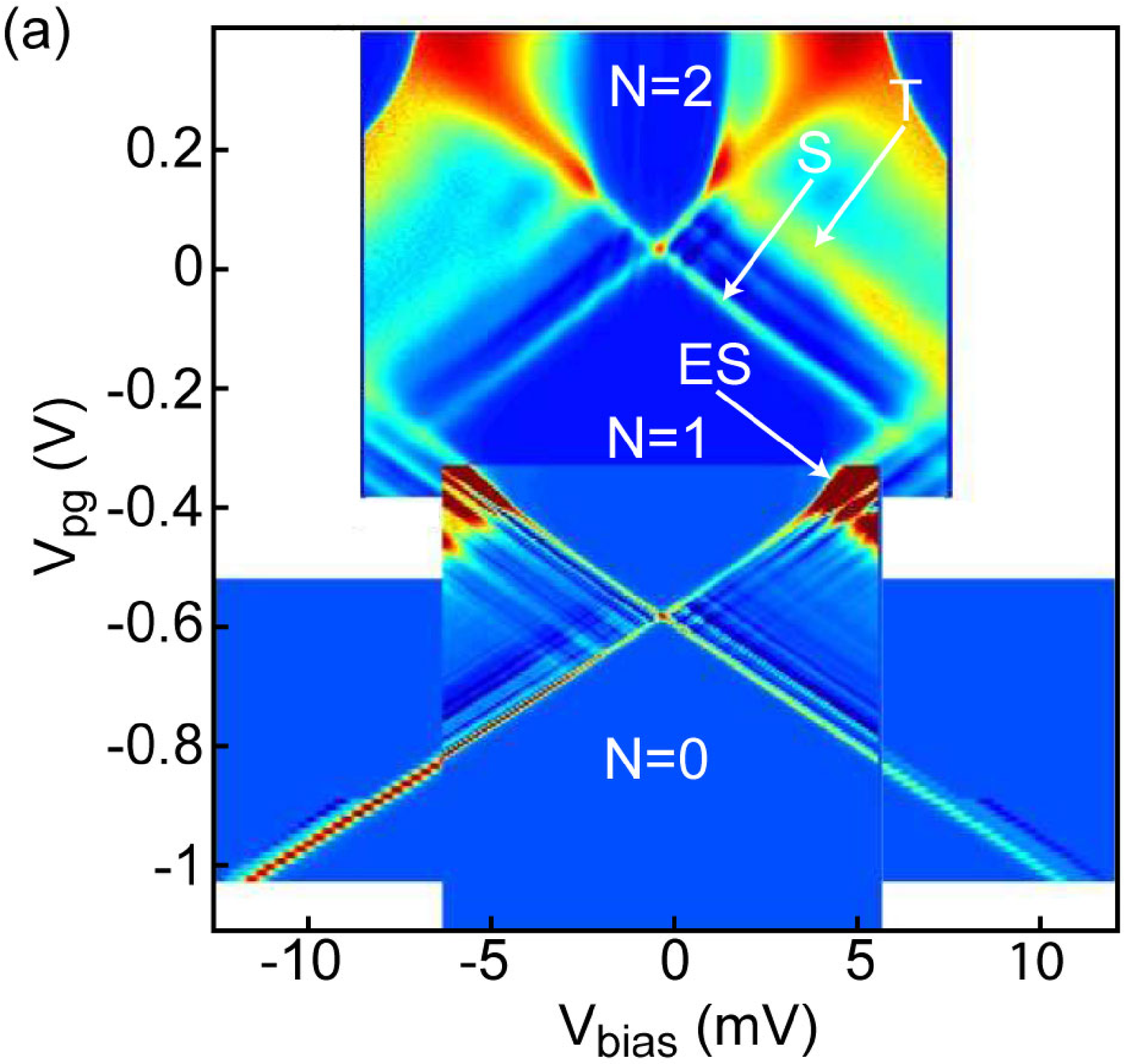}}\hfill
\parbox[b]{6cm}{
\includegraphics[width=6cm]{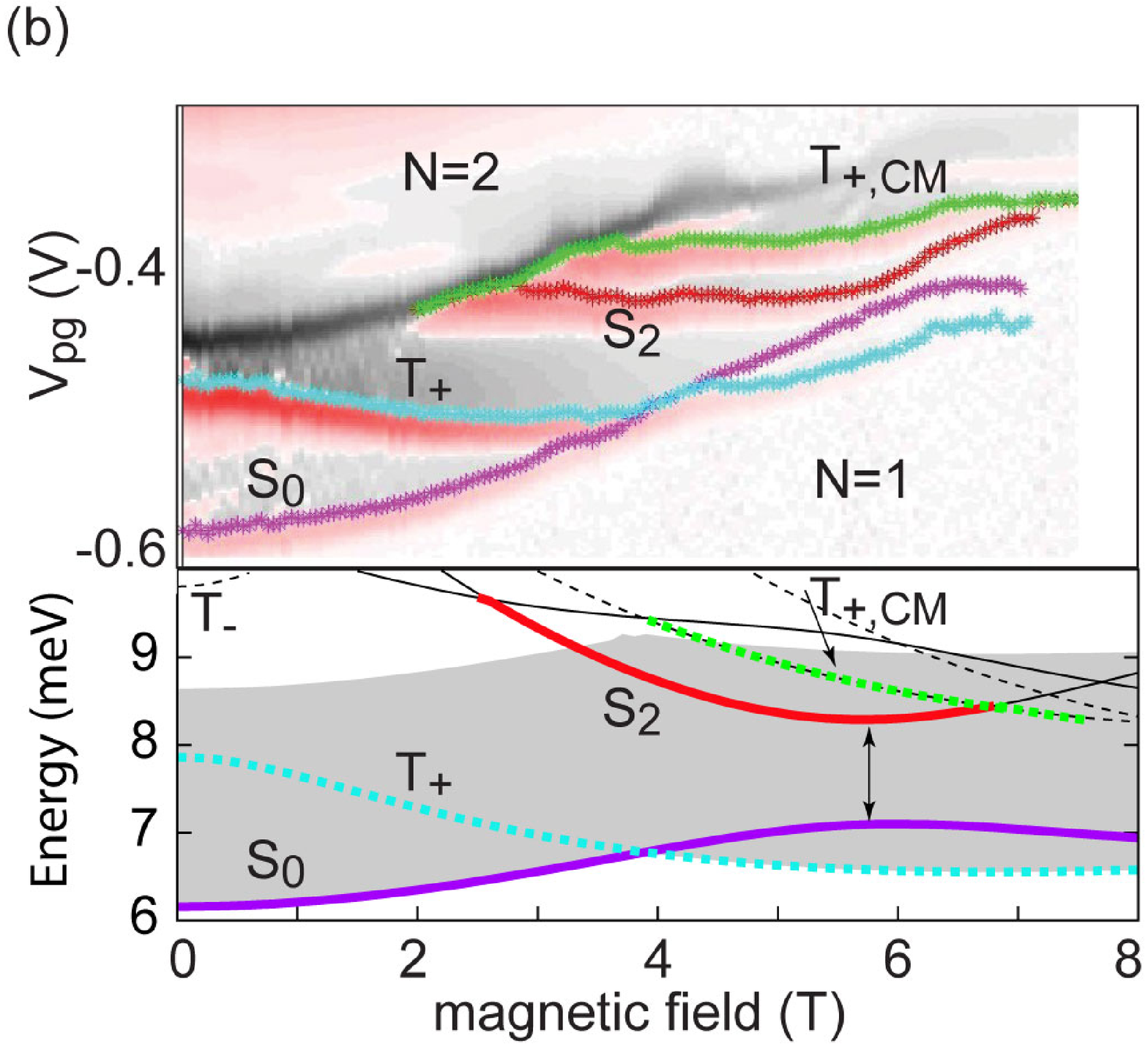}}
\caption{(a) Differential conductance of the few-electron quantum dot as a function of the plunger gate voltage $V_\mathrm{pg}$ and the source--drain voltage $V_\mathrm{bias}$. Electron numbers are indicated. ES labels the resonance of the first excited state of the 1-electron dot, S and T label the singlet- and triplet state transitions of the 2-electron system. (b) The top panel shows the measured excitation spectrum of QD He in a magnetic field. The bottom panel is the corresponding energy spectrum of QD He calculated with EXD. The gray-shaded region indicates the bias-window of $V_\mathrm{bias}=2.5$~meV of the experiment.}
\label{fig3}
\end{figure}
Figure~\ref{fig3} shows the resulting Coulomb-blockade diamonds with the electron numbers $N$ indicated. The QD has a charging energy $\Delta\mu_2=6.9$~meV for the second electron. The lowest excited state ES of the 1-electron system has an energy $\Delta_1=5$~meV above the ground state. Therefore, confinement and interaction effects are of comparable order of magnitude in this system. Singlet (S) and Triplet (T) resonances are observed in the differential conductance of the 2-electron dot. The singlet--triplet separation $J$ is about 2~meV.

For the investigation of the excitation spectrum of quantum dot Helium, i.e., the two-electron system, in a magnetic field, the QD was tuned into a slightly more open regime allowing the plunger gate to cover the range from $N=1$ to 3. The excitation spectrum of the system was then recorded at fixed source--drain voltage $V_\mathrm{bias}=2.5$~mV as a function of plunger gate voltage $V_\mathrm{pg}$ and magnetic field $B$.
The result\cite{Ellenberger06} is shown in Fig.~\ref{fig3}(b) (upper panel).
A number of transitions can be followed as a function of magnetic field: most prominent is the transition from the 1-electron ground state to the 2-electron singlet state, labeled $S_0$. The first excitation at low $B$ is the triplet state labeled $T_+$ which becomes the ground state at a magnetic field $B\approx 4$~T. The transition labeled $S_2$ appears at $B\approx 3$~T in the bias window and shows an avoided crossing with $S_0$ indicating its spin-singlet character. The state $T_\mathrm{+,CM}$ appearing at even higher $B$ can only be identified by comparing to the results of an exact diagonalization calculation (EXD) to be described below. It turns out to be combined excitation of the center of mass (CM) motion, the relative motion (+) and the spin degree-of-freedom (T).

\section{Theoretical Methods and Calculations for Quantum Dot Helium} 

In order to interpret the measured excitation spectra in detail, we present an exact diagonalization (EXD) and an approximate
(generalized Heitler-London, GHL) microscopic treatment for two electrons in a
{\it single\/} elliptic QD specified by parameters that correspond to our
experimental device.\cite{Ellenberger06}

The Hamiltonian for the two 2D interacting electrons is given by 
\begin{equation}
{\cal H} = H({\bf r}_1)+H({\bf r}_2)+ \gamma e^2/(\kappa r_{12}),
\label{ham}
\end{equation}
where the last term is the Coulomb repulsion, $\kappa$ (12.5 for GaAs) is the
dielectric constant, and $r_{12} = |{\bf r}_1 - {\bf r}_2|$. The prefactor
$\gamma$ accounts for the reduction of the Coulomb strength due to the finite
thickness of the electron layer in the $z$ direction and for any additional
screening effects due to the gate electrons. $H({\bf r})$ is the
single-particle Hamiltonian for an electron in an external perpendicular
magnetic field ${\bf B}$ and an appropriate confinement potential.
For an elliptic QD, the single-particle Hamiltonian is written as
\begin{equation}
H({\bf r}) = T + \frac{1}{2} m^* (\omega^2_{x} x^2 + \omega^2_{y} y^2),
\label{hsp}
\end{equation}
where $T=({\bf p}-e{\bf A}/c)^2/2m^*$, with ${\bf A}=0.5(-By,Bx,0)$ being the
vector potential in the symmetric gauge. The effective mass is $m^*=0.07m_0$,
and ${\bf p}$ is the linear momentum of the electron. The second term is the
external confining potential. In the Hamiltonian (\ref{hsp}), we neglect the
Zeeman contribution due to the negligible value ($g^* \approx 0$) of the 
effective Land\'{e} factor in our sample.

\subsection{Generalized Heitler--London Approach}

The GHL method for solving the Hamiltoninian (\ref{ham}) consists of two steps.
In the first step, we solve self-consistently the ensuing
unrestricted Hartree-Fock (UHF) equations allowing for lifting of the
double-occupancy requirement (imposing this requirement gives the
{\it restricted\/} HF method, RHF).
For the $S_z=0$ solution, this step produces two single-electron
orbitals $u_{L,R}({\bf r})$ that are localized left $(L)$ and right $(R)$ of the
center of the QD [unlike the RHF method that gives a single doubly-occupied
elliptic (and symmetric about the origin) orbital].
At this step, the many-body wave function is a single Slater
determinant $\Psi_{\mathrm{UHF}} (1\uparrow,2\downarrow) \equiv
| u_L(1\uparrow)u_R(2\downarrow) \rangle$ made out of the two occupied UHF
spin-orbitals $u_L(1\uparrow) \equiv u_L({\bf r}_1)\alpha(1)$ and
$u_R(2\downarrow) \equiv u_R({\bf r}_2) \beta(2)$, where
$\alpha (\beta)$ denotes the up (down) [$\uparrow (\downarrow)$] spin.
This UHF determinant is an eigenfunction of the projection $S_z$ of the total
spin $\hat{S} = \hat{s}_1 + \hat{s}_2$, but not of $\hat{S}^2$ (or the parity
space-reflection operator).

In the second step, we restore the broken parity and total-spin symmetries by
applying to the UHF determinant the projection operator\cite{yyl8,yl3}
${\cal P}_{\rm spin}^{s,t}=1 \mp \varpi_{12}$, where the
operator $\varpi_{12}$ interchanges the spins of the two electrons;
the upper (minus) sign corresponds to the singlet.
The final result is a generalized Heitler-London two-electron wave function
$\Psi^{s,t}_{\mathrm{GHL}} ({\bf r}_1, {\bf r}_2)$ for the ground-state singlet
(index $s$) and first-excited triplet (index $t$), which uses
the UHF localized orbitals,
\begin{equation}
\Psi^{s,t}_{\mathrm{GHL}} ({\bf r}_1, {\bf r}_2) \propto
{\bf (} u_L({\bf r}_1) u_R({\bf r}_2) \pm u_L({\bf r}_2) u_R({\bf r}_1) {\bf )}
\chi^{s,t},
\label{wfghl}
\end{equation}
where $\chi^{s,t} = (\alpha(1) \beta(2) \mp \alpha(2) \beta(1))$ is the spin
function for the 2$e$ singlet and triplet states.
The general formalism of the 2D UHF equations and of the subsequent restoration
of broken spin symmetries can be found in Refs.\ \cite{yyl8,yyl9,yl1,yl3}.

The use of {\it optimized\/} UHF orbitals in the GHL is suitable for treating
{\it single elongated\/} QDs. The GHL is equally applicable to double QDs with
arbitrary interdot-tunneling coupling.\cite{yyl8,yl3} In contrast,
the Heitler-London (HL) treatment\cite{hl} (known also as Valence bond),
where non-optimized ``atomic'' orbitals of two isolated QDs are used, is
appropriate only for the case of a double dot with small interdot-tunneling
coupling.\cite{burk}

The orbitals $u_{L,R}({\bf r})$ are expanded in a real Cartesian
harmonic-oscillator basis, i.e.,
\begin{equation}
u_{L,R}({\bf r}) = \sum_{j=1}^K C_j^{L,R} \varphi_j ({\bf r}),
\label{uexp}
\end{equation}
where the index $j \equiv (m,n)$ and $\varphi_j ({\bf r}) = X_m(x) Y_n(y)$,
with $X_m(Y_n)$ being the eigenfunctions of the one-dimensional oscillator in the
$x$($y$) direction with frequency $\omega_x$($\omega_y$). The parity operator
${\cal P}$ yields ${\cal P} X_m(x) = (-1)^m X_m(x)$, and similarly for $Y_n(y)$.
The expansion coefficients $C_j^{L,R}$ are real for $B=0$ and complex for finite
$B$. In the calculations we use $K=54$ and/or $K=79$, yielding convergent 
results.

\subsection{Exact Diagonalization}

In the EXD method, the many-body wave function is written as a linear
superposition over the basis of non-interacting two-electron determinants, i.e.,
\begin{equation}
\Psi^{s,t}_{\mathrm{EXD}} ({\bf r}_1, {\bf r}_2) =
\sum_{i < j}^{2K} \Omega_{ij}^{s,t} | \psi(1;i) \psi(2;j)\rangle,
\label{wfexd}
\end{equation}
where $\psi(1;i) = \varphi_i(1 \uparrow)$ if $1 \leq i \leq K$ and
$\psi(1;i) = \varphi_{i-K}(1 \downarrow)$ if $K+1 \leq i \leq 2K$ [and
similarly for $\psi(2,j)$].
The total energies $E^{s,t}_{\mathrm{EXD}}$ and the coefficients
$\Omega_{ij}^{s,t}$ are obtained through a ``brute force'' diagonalization of
the matrix eigenvalue equation corresponding to the Hamiltonian in Eq.\
(\ref{ham}). The EXD wave function does not
immediately reveal any particular form, although, our calculations below
show that it can be approximated by a GHL wave function in the case of the
elliptic dot under consideration.

\subsection{Results and Comparison with Measurements}

To model the experimental quantum dot device, we take, following Ref.~1, $\hbar \omega_x=4.23$ meV, $\hbar \omega_y=5.84$ meV, and
$\gamma=0.862$. The corresponding anisotropy is $\omega_y/\omega_x=1.38$,
indicating that the quantum dot considered here is closer to being circular
than in other experimental systems.\cite{Kyriakidis02,Zumbuhl04} 

\begin{figure}[t]
\parbox[b]{6cm}{
\includegraphics[width=6.0cm]{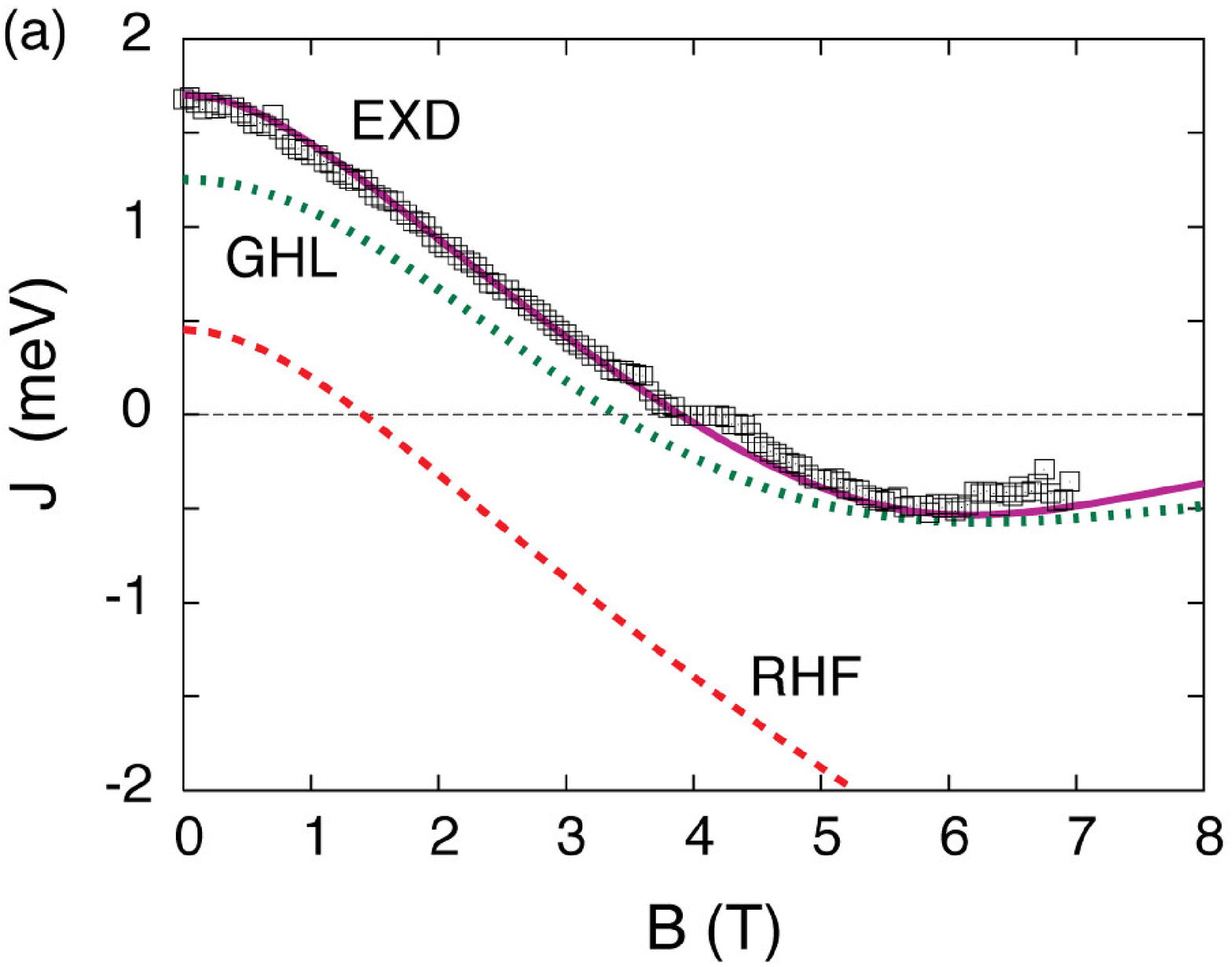}}\hfill
\parbox[b]{6cm}{
\includegraphics[width=6.0cm]{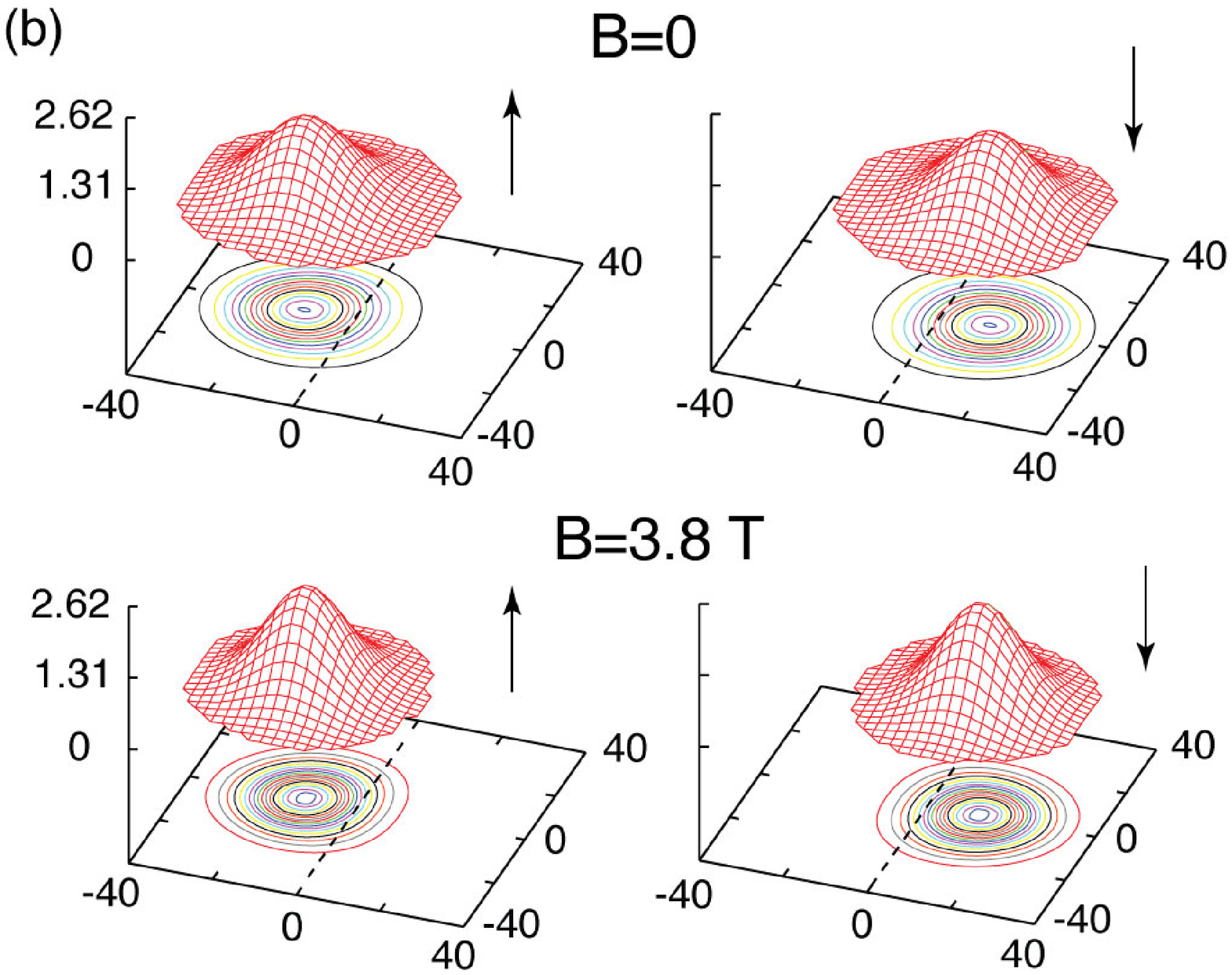}}
\caption{(a) Comparison of $J(B)$ calculated with different
methods and the experimental results (open squares). Solid line:
EXD. Dotted line: GHL. Dashed line: RHF.
(b) Single-particle UHF orbitals (modulus square) that are used in the 
construction of the GHL wave function in Eq.\ (\ref{wfghl}).
Lengths in nm and orbital densities in 10$^{-3}$ nm$^{-2}$. Arrows indicate up and down spins. For the parameters used in the calculation to model the anisotropic QD,
see text.}
\label{fig1_th_hmfw}
\end{figure}

As shown in Ref.~1 and Fig.~\ref{fig3}(b), the experimental findings can be 
quantitatively interpreted by comparing to the results of the EXD calculations 
for two electrons in an anisotropic harmonic confinement potential with the
parameters listed above. All the states observed in the measured spectra
(as a function of the magnetic field) can be unambiguously identified\cite{Ellenberger06} with calculated ground-state and excited states of the 
two-electron Hamiltonian [see Fig.~\ref{fig3}(b)]. In particular, the calculated magnetic-field-dependent 
energy splitting, $J_{\mathrm{EXD}}(B)=E^t_{\mathrm{EXD}}(B)-E^s_{\mathrm{EXD}}(B)$,
between the two lowest singlet ($S_0$) and triplet ($T_+$) states is 
found to be in remarkable agreement with the experiment [see Fig.\ 
\ref{fig1_th_hmfw}(a)]. 

A deeper understanding of the structure of the many-body wave function can
be acquired by comparing the measured $J(B)$ with that calculated within the 
GHL and RHF approximations. To facilitate the comparisons, the calculated 
$J_{\mathrm{GHL}}(B)$ and $J_{\mathrm{RHF}}(B)$ curves are plotted also in Fig.~\ref{fig1_th_hmfw}(a),
along with the EXD result and the measurements. Both the RHF and GHL schemes 
are appealing intuitively, because they minimize the total energy using 
single-particle orbitals. It is evident, however, from Fig.\ \ref{fig1_th_hmfw}(a) 
that the RHF method, which assumes that both electrons occupy a common 
single-particle orbital, is not able to reproduce the experimental findings. 
On the contrary, the GHL approach, which allows the two electrons to occupy two 
spatially separated orbitals, appears to be a good approximation. 
Plotting the two GHL orbitals [see Fig.\ \ref{fig1_th_hmfw}(b)] for the singlet 
state clearly demonstrates that the two electrons do not occupy the same orbital,
but rather fill states that are spatially separated significantly. 


The UHF orbitals from which the GHL singlet state is constructed
[see Eq.\ (\ref{wfghl})] are displayed on Fig.\ \ref{fig1_th_hmfw}(b) for 
both the $B=0$ and $B=3.8$ T cases. The spatial shrinking of these orbitals 
at the higher $B$-value illustrates the ``dissociation'' of the electron dimer 
with increasing magnetic field. The asymptotic convergence (beyond the ST point)
of the energies of the singlet and triplet states, [i.e., $J(B) \rightarrow 0$ 
as $B \rightarrow \infty$] is a reflection of the dissociation of the 2$e$ 
molecule, since the ground-state energy of two fully spatially separated 
electrons (zero overlap) does not depend on the total spin. We stress again
that the RHF, which corresponds to the more familiar physical picture of a
QD-Helium atom, fails to describe this dissociation, because $J_{\mathrm{RHF}}(B)$
diverges as the value of the magnetic field increases.

\begin{figure}[t]
\centering
\includegraphics[width=6.0cm]{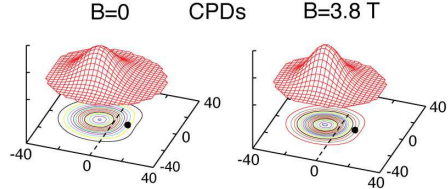}
\caption{CPDs extracted from the EXD wave function for the singlet state for $B=0$
and $B=3.8$ T. The CPD expresses the conditional probability for finding the
second electron at position ${\bf r}$ given that the first electron is 
located at ${\bf r}_0$ (denoted by a heavy solid dot).
For the parameters used in the calculation to model the anisotropic QD,
see text. Lengths in nm and CPDs in arbitrary units.}
\label{fig3_th_hmfw}
\end{figure}

In contrast to the RHF, the GHL wave function is able to capture the importance
of correlation effects. Further insight into the importance of correlations
in our QD device can be gained through inspection\cite{Ellenberger06} of the 
conditional probability distributions\cite{yl4} (CPDs) associated with
the EXD solutions; see an illustration in Fig.\ \ref{fig3_th_hmfw}.
Indeed, already at zero magnetic field, the calculated CPDs 
provide further support of the physical picture of two localized electrons 
forming a state resembling an H$_2$-type\cite{yyl8,Ellenberger06} Wigner 
molecule.\cite{yl5,Egger99}

\subsection{Degree of Entanglement}

Further connections between the strong correlations found in our microscopic
treatment and the theory of quantum computing\cite{burk} can be made
through specification of the degree of entanglement between the two
localized electrons in the molecular dimer. For two electrons, we can
quantify the degree of entanglement by calculating a well-known measure
of entanglement such as the von Neumann entropy\cite{you,yl6} for {\em indistinguishable} particles.
To this end, one needs to bring the EXD wave function into a diagonal form 
(the socalled ``canonical form''\cite{you,schl2}), i.e.,
\begin{equation}
\Psi^{s,t}_{\mathrm{EXD}} ({\bf r}_1, {\bf r}_2) =
\sum_{k=1}^M z^{s,t}_k | \Phi(1;2k-1) \Phi(2;2k) \rangle,
\label{cano}
\end{equation}
with the $\Phi(i)$'s being appropriate spin orbitals resulting from a unitary
transformation of the basis spin orbitals $\psi(j)$'s [see Eq.\ (\ref{wfexd})];
only terms with $z_k \neq 0$ contribute. The upper bound $M$ can be
smaller (but not larger) than $K$ (the dimension of the
single-particle basis); $M$ is referred to as the Slater rank.
One obtains the coefficients of the canonical expansion from the fact that
the $|z_k|^2$ are eigenvalues of the hermitian matrix $\Omega^\dagger \Omega$
[$\Omega$, see Eq.\ (\ref{wfexd}), is antisymmetric]. The von Neumann
entropy is given by ${\cal S} = -\sum_{k=1}^M |z_k|^2 \log_2(|z_k|^2)$ with the
normalization $\sum_{k=1}^M |z_k|^2 =1$.

\begin{figure}[t]
\centering\includegraphics[width=5cm]{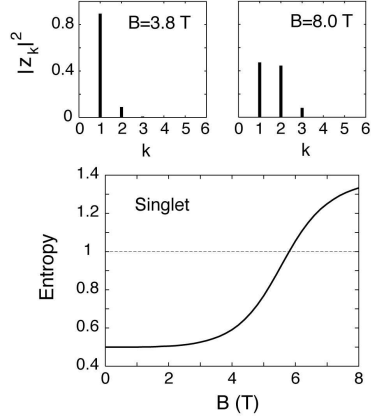}
\caption{
Von Neumann entropy for the lowest singlet EXD state of the elliptic dot as a 
function of the magnetic field $B$. 
On the top, we show histograms for the $|z_k|^2$
coefficients [see Eq.\ (\ref{cano})] of the singlet state at $B=3.8$ T (left)
and $B=8.0$ T (right) illustrating the dominance of two determinantal 
configurations (in agreement with the GHL picture). Note the small third
coefficient $|z_3|^2=0.081$ for $B=8.0$ T.
For the parameters used to model our device, see text.
}
\label{fig4_th_hmfw}
\end{figure}

The EXD singlet has obviously a Slater rank $M > 2$.
The von Neumann entropy for the EXD singlet (${\cal S}^s_{\mathrm{EXD}}$) is
displayed in Fig.\ \ref{fig4_th_hmfw}. It is remarkable that  
${\cal S}^s_{\mathrm{EXD}}$ increases with increasing $B$, but remains close to 
unity for large $B$, although the 
maximum allowed mathematical value is $\log_2(K)$ [for example for
$K=79$, $\log_2(79)=6.3$]. The saturation
of the entropy for large $B$ to a value close to unity reflects the dominant
(and roughly equal at large $B$) weight of two configurations in the canonical 
expansion [see Eq.\ (\ref{cano})] of the EXD wave function, which are related\cite{yl6} to the two terms in the canonical expansion of the GHL singlet. This 
is illustrated by the histograms of the $|z_k^s|^2$ coefficients for $B=3.8$ T 
and $B= 8.0$ T in Fig.\  \ref{fig4_th_hmfw} (top). Notice that the ratio
$|z_2|^2/|z_1|^2$ reflects the extent of the overlap between the two GHL
orbitals\cite{yl6}, with the ratio increasing for smaller overlaps
(corresponding to a more complete dissociation of the Wigner molecule).

The above discussion illustrates that microscopic calculations that are
shown to reproduce experimental spectra\cite{Ellenberger06} can be used
to extract valuable information that allows assessment of the suitability
of a given device for quantum computations.

\section{Conclusions}

Measurements and theoretical interpretation were presented of the magnetic field dependent excitation spectra of a two-electron quantum dot. The quantum dot is based on an Al$_x$Ga$_{1-x}$As parabolic quantum well with effective $g^\star$-factor close to zero. Results of tunneling spectroscopy of the four lowest states were compared to exact diagonalization calculations and a generalized Heitler--London approximation and good agreement was found. Electronic correlations, associated with the formation of an H$_2$-type Wigner molecule, turn out to be very important in this system.

\section*{Acknowledgements}

We gratefully acknowledge financial support from the Schweizerischer Nationalfonds, the US D.O.E. (Grant No. FG05-86ER45234), and the NSF (Grant No. DMR-0205328).


\section*{References}

\begin{thebibliography}{100}

\bibitem{Ellenberger06}
C. Ellenberger, T.Ihn, C. Yannouleas, U. Landman,
K. Ensslin, D. Driscoll, and A.C. Gossard,
Phys. Rev. Lett. {\bf 96}, 126806 (2006).
\bibitem{Kouwenhoven97} L.P. Kouwenhoven, T.H. Oosterkamp, M.W.S. Danoesastro, M. Eto, D.G. Austing, T. Honda, S. Tarucha, Science {\bf 278}, 1788 (1997).
\bibitem{Shayegan88} M. Shayegan, T. Sajoto, M. Santos, C. Silvestre, Appl. Phys. Lett. {\bf 53}, 791 (1988); E.G. Gwinn, R.M. Westervelt, P.F. Hopkins, A.J. Rimberg, M. Sundaram, A.C. Gossard, Phys. Rev. B {\bf 39}, 6260 (1989).
\bibitem{Salis97} G. Salis, B. Graf, K. Ensslin, K. Campman, K. Maranowski, A.C. Gossard, Phys. Rev. Lett. {\bf 79}, 5106 (1997).
\bibitem{Salis01} G. Salis, Y. Kato, K. Ensslin, D.C. Driscoll, A.C. Gossard, D.D. Awshalom, Nature {\bf 414}, 619 (2001). 
\bibitem{Salis99} G. Salis, T. Heinzel, K. Ensslin, O. Homan, W. Bachtold, K. Maranowski, A.C. Gossard, Microelectronic Engineering {\bf 47}, 175 (1999).
\bibitem{Lindemann02} S. Lindemann, T. Ihn, T. Heinzel, K. Ensslin, K. Maranowski, A.C. Gossard, Physica E {\bf 13}, 638 (2002).
\bibitem{Lindemann02a} S. Lindemann, T. Ihn, T. Heinzel, W. Zwerger, K. Ensslin, K. Maranowski, A.C. Gossard, Phys. Rev. B {\bf 66}, 195314 (2002).
\bibitem{Loss98} D. Loss and D. P. Di Vincenzo, Phys. Rev. A {\bf 57}, 120 (1998). 
\bibitem{Cerletti05} V. Cerletti, W.A. Coish, O. Gywatt, D. Loss, Nanotechnology {\bf 16}, R27 (2005).
\bibitem{Field93} M. Field, C.G. Smith, M. Pepper, D.A. Ritchie, J.E.F. Frost, G.A.C. Jones, D.G. Hasko, Phys. Rev. Lett. {\bf 93}, 1311 (1993).
\bibitem{yyl8}
C. Yannouleas and U. Landman,
Int. J. Quantum Chem. {\bf 90}, 699 (2002).
\bibitem{yyl9}
C. Yannouleas and U. Landman,
J. Phys.: Condens. Matter {\bf 14}, L591 (2002).
\bibitem{yl1}
C. Yannouleas and U. Landman,
Phys. Rev. B {\bf 68}, 035325 (2003).
\bibitem{yl3}
C. Yannouleas and U. Landman,
Eur. Phys. J. D {\bf 16}, 373 (2001).
\bibitem{hl}
H. Heitler and F. London,
Z. Phys. {\bf 44}, 455 (1927).
\bibitem{burk}
G. Burkard, D. Loss, and D.P. DiVincenzo,
Phys. Rev. B {\bf 59}, 2070 (1999).
\bibitem{Kyriakidis02}
J. Kyriakidis, M. Pioro-Ladriere, M. Ciorga, A.S. Sachrajda, P. Hawrylak,
Phys. Rev. B {\bf 66}, 035320 (2002).
\bibitem{Zumbuhl04}
D.M. Zumb\"{u}hl, C.M. Marcus, M.P. Hanson, A.C. Gossard, 
Phys, Rev. Lett. {\bf 93}, 256801 (2004); 
C. Yannouleas and U. Landman, Proc. Natl. Acad. Sci. (USA) {\bf 103},
10600 (2006).
\bibitem{yl4}
C. Yannouleas and U. Landman,
Phys. Rev. Lett. {\bf 85}, 1726 (2000);
Phys. Rev. B {\bf 70}, 235319 (2004);
P.A. Maksym,
Phys. Rev. B {\bf 53}, 10871 (1996).
\bibitem{yl5}
C. Yannouleas and U. Landman,
Phys. Rev. Lett. {\bf 82}, 5325 (1999).
\bibitem{Egger99}
R. Egger, W. H\"ausler, C.H. Mak, and H. Grabert,
Phys. Rev. Lett. {\bf 82}, 3320 (1999).
\bibitem{you}
R. Paskauskas and L. You,
Phys. Rev. A {\bf 64}, 042310 (2001).
\bibitem{yl6}
C. Yannouleas and U. Landman,
phys. stat. sol. (a) {\bf 203}, 1160 (2006).
\bibitem{schl2}
J. Schliemann, J. Cirac, M. Kus, M. Lewenstein, and D. Loss,
Phys. Rev. A {\bf 64}, 022303 (2001).
\end{thebibliography}
\end{document}